\documentclass[a4paper,11pt]{article}
\usepackage{geometry}
\usepackage[T1]{fontenc}
\usepackage[utf8]{inputenc}
\usepackage[english]{babel}
\usepackage{lmodern}
\usepackage{graphicx,subcaption}
\usepackage{color}
\usepackage{url}
\usepackage{epsf,epsfig}
\usepackage{amsmath,amsfonts,amssymb}
\usepackage[colorlinks=true, urlcolor=blue]{hyperref}
\everymath{\displaystyle} 
\usepackage{authblk}
\newcommand{\be}{\begin{equation}}
\newcommand{\ee}{\end{equation}}
\newcommand{\bi}{\begin{itemize}}
\newcommand{\ei}{\end{itemize}}
\newcommand{\ba}{\begin{array}}
\newcommand{\ea}{\end{array}}
\newcommand{\bea}{\begin{eqnarray}}
\newcommand{\eea}{\end{eqnarray}}
\newcommand{\nn}{\nonumber}

\title{\textbf{$\pi^0, \eta, \eta^\prime$ two-photon transition form factors \\ in the holographic soft-wall model \\ and  contributions to $(g-2)_\mu$}}
\author[1]{Pietro Colangelo\thanks{\href{{mailto:pietro.colangelo@ba.infn.it}}{pietro.colangelo@ba.infn.it}}}
\author[1]{Floriana Giannuzzi\thanks{\href{{mailto:floriana.giannuzzi@ba.infn.it}}{floriana.giannuzzi@ba.infn.it}}}
\author[1]{Stefano Nicotri\thanks{\href{mailto:nicotri@infn.it}{nicotri@infn.it}}}
\affil[1]{\small \emph{INFN -- Istituto Nazionale di Fisica Nucleare -- Sezione di Bari} \protect\\ \emph{Via Orabona 4, 70125, Bari, Italy}}
\date{}

\begin{document}
\begin{flushright}{BARI-TH/23-744} \end{flushright}
{\let\newpage\relax\maketitle}
\maketitle

\begin{abstract}
 We compute the two-photon transition form factors of the light pseudoscalar $\pi^0$, $\eta$ and $\eta^\prime$ mesons in a soft-wall holographic model of QCD with a solution of the $U(1)_A$ problem. We compare the results with the experimental data in different ranges of photon virtualities. The obtained transition form factors are used to determine the $\pi^0$, $\eta$ and $\eta^\prime$ pole terms in the hadronic light-by-light scattering contribution to the anomalous magnetic moment of the muon.
\end{abstract}
 
\section{Introduction}
The aim of the present study is twofold. 
The first purpose is to exploit a model proposed by two of us to face the $U(1)_A$ problem of quantum chromodynamics using the holographic approach \cite{Giannuzzi:2021euy}, within the class of the so-called AdS/QCD soft-wall models \cite{Karch:2006pv}. 
In particular, we compute the transition form factors of light pseudoscalar mesons to two photons in a wide range of photon virtualities. 
The comparison with experiment will allow us to shed light on the successful features and on the drawbacks of the model, and on the possible improvements. 
The second (related) purpose is to compute the light pseudoscalar meson contributions to the anomalous magnetic moment of the muon, i.e. their pole contribution to the hadronic light-by-light (HLbL) scattering amplitude.\footnote{For an introduction and references to the early studies see \cite{Melnikov:2006sr,Jegerlehner:2017gek}. }
The recently confirmed tension between the measurement \cite{Muong-2:2006rrc,Muong-2:2021ojo} and the Standard Model (SM) prediction for $a_\mu=(g-2)_\mu / 2$ \cite{Aoyama:2020ynm} motivates the efforts for further scrutinizing the SM expectation. 
The HLbL pole contribution of the various hadrons represents a fraction of the total hadronic contribution dominated by the hadronic vacuum polarization (HVP), as discussed in the White Paper of the Muon $(g-2)$ Theory Initiative \cite{Aoyama:2020ynm}, however it is affected by a sizeable error. 
The contribution of light pseudoscalar mesons, after past analyses in holographic approaches which also considered axial-vector and scalar mesons \cite{Hong:2009zw,Cappiello:2010uy,Cappiello:2019hwh,Leutgeb:2019gbz,Leutgeb:2019zpq,Cappiello:2021vzi,Leutgeb:2021mpu}, has been recently evaluated in an updated model \cite{Leutgeb:2022lqw}: the comparison with the results obtained here will strenghten the confidence on the precision achieved by such models.\footnote{A general discussion on the HLbL pole contribuition to $a_\mu$ and references can be found in \cite{Aoyama:2020ynm}.}
This is an important issue, due to the role played by the lepton anomalous magnetic moments as powerful tests of the SM.
A remarkable feature of holographic models is that the Melnikov-Vainshtein longitudinal short-distance constraint on the HLbL four-photon amplitude \cite{Melnikov:2003xd} can be fulfilled by the summation of contributions from the infinite tower of axial-vector mesons \cite{Leutgeb:2019gbz}.
 
To compute the two-photon transition form factors of $\pi^0$, $\eta$ and $\eta^\prime$ we adopt a holographic model encoding a solution of the $U(1)_A$ problem in QCD. 
The QCD Lagrangian ${\cal L}_{QCD}$ with $n_f$ massless quarks exhibits a global $U(n_f)_R\times U(n_f)_L$ flavour symmetry, spontaneously broken to $SU(n_f)_V\times U(1)_V \times U(1)_A $ with $n_f^2-1$ Goldstone bosons in the low energy spectrum. $SU(n_f)_V$ is conserved, $U(1)_V$ is the baryon number conservation. 
The Weinberg $U(1)$ problem \cite{Weinberg:1975ui} is the lack of evidence of the $U(1)_A$ symmetry under quark field transformations
\begin{equation}
 q_i(x) \to e^{i \theta \gamma_5}q_i(x) , \hspace*{1cm} \bar q_i(x) \to \bar q_i(x) e^{i \theta \gamma_5} , \label{eq:chir}
\end{equation} 
and the absence of an additional Goldstone boson in the spectrum signaling the spontaneous breaking of such a symmetry: the $\eta^\prime$ meson is much heavier than $m_{\eta^\prime}=\sqrt{3} m_\pi$ expected in that case by chiral theory.
Remarkably, $U(1)_A$ is anomalous: it is broken by quantum corrections. 
The singlet axial current satisfies the anomaly condition
\begin{equation}
 \partial_\mu J^\mu_A=-\frac{g_s^2}{32 \pi^2} G_{\mu \nu}^a \tilde G^{a \mu \nu} \,\, \label{eq:anomaly}
\end{equation}
for massless quarks, with $G_{\mu \nu}^a$ the gluon field strength, $ \tilde G_{\mu \nu}^a=\frac{1}{2}\epsilon_{\mu \nu \alpha \beta}G^{a \alpha \beta}$ the dual strength, $a$ the color index. 
In the path integral the Jacobian of the transformations \eqref{eq:chir} is not 1, and a chiral rotation modifies the QCD Lagrangian by the term $\theta \frac{g_s^2}{32 \pi^2} G^{a\mu \nu} \tilde G^a_{\mu \nu}$. 
This implies that the CP-odd term ${\cal L}_{\theta}= \bar \theta \frac{g_s^2}{32 \pi^2} G_{\mu \nu}^a \tilde G^{a \mu \nu} $ must be added to ${\cal L}_{QCD}$ ($\bar \theta$ being the effective parameter) with dynamical consequences at low energy, recognized using effective Lagrangians \cite{DiVecchia:1979pzw,DiVecchia:1980yfw,Witten:1980sp,Kawarabayashi:1980dp} and lattice QCD computations \cite{Vicari:2008jw,Bonati:2015vqz,Lombardo:2020bvn}.
Since $P$ and $T$ violations are not observed in strong interactions (the bound $|\bar \theta| < 3 \times 10^{-10}$ results from the experimental higher bound on the neutron electric dipole moment \cite{ParticleDataGroup:2022pth}) the problem of explaining the tiny value of $\bar \theta$ arises, with possible solutions involving axions from different mechanisms \cite{Peccei:1977hh,Shifman:1979if,Zhitnitsky:1980tq,Dine:1981rt}.

 The identification of a $U(1)_A$ breaking mechanism was provided by the t'Hooft recognition of nontrivial topological gauge field configurations, the instantons, with action $S$ proportional to $\int \mathrm{d}^4x \, \mathrm{Tr}[ G^{\mu \nu} \tilde G_{\mu \nu}]$, contributing to the path integral via the anomaly \cite{tHooft:1976rip}. 
 However, the topological term is a total derivative, it does not contribute to any order in perturbation theory. 
 Its contribution, of order $\exp(-1/g_s^2)$, can be sizeable only for large values of the strong coupling constant, in the nonperturbative sector of QCD. 
 Lattice QCD analyses, overcoming the difficulty to simulate such a term, give results for the $\eta^\prime$ mass close to the experimental value from the two-point correlation function of the $G_{\mu \nu} \tilde G^{\mu \nu}$ operator.\footnote{For a discussion and references to previous studies see \cite{Bali:2021qem}.}
 
A method to investigate the nonperturbative sector of QCD is based on the holographic correspondence \cite{Maldacena:1997re,  Witten:1998qj, Gubser:1998bc}. 
Bottom-up holographic models of QCD have been constructed, setting up a correspondence between bulk fields in an asymptotically Anti-de Sitter (AdS) five-dimensional space and local QCD operators on the AdS$_5$ boundary \cite{Erlich:2005qh,Karch:2006pv}. 
A soft-wall model addressing the $U(1)_A$ problem has been developed in \cite{Giannuzzi:2021euy}, with a calculation of the $\eta^\prime$ mass and of the QCD topological susceptibility. 
Here, we focus on the two-photon transition form factors of the light pseudoscalar mesons. 
Other holographic approaches facing the $U(1)_A$ problem are described in \cite{Katz:2007tf,Arean:2016hcs,Bigazzi:2015bna}.

The plan of the paper is as follows. 
Section~\ref{sec:model} contains a description of the main features of the soft-wall holographic model encoding a solution to the $U(1)_A$ problem. 
The two-photon transition form factors of $\pi^0$, $\eta$ and $\eta^\prime$ are computed in the model in section~\ref{sec:TFF}, and are compared to the experimental measurements. 
In section~\ref{sec:g-2} the HLbL pole contribution of $\pi^0$, $\eta$, $\eta^\prime$ and $\pi^0(2S)$ to $a_\mu$ is determined. 
The conclusions and the perspectives for improvements are presented in the last section.

\section{Soft-wall holographic model of QCD and the $U(1)_A$ problem}\label{sec:model}

The soft-wall holographic model of QCD is defined in the $5d$ space with background AdS geometry, with line element
\begin{equation}
\mathrm{d}s^2=g_{MN}\mathrm{d}x^M \mathrm{d}x^N=\frac{R^2}{z^2} \left(\eta_{\mu \nu} \mathrm{d}x^\mu \mathrm{d}x^\nu - \mathrm{d}z^2\right). 
\end{equation}
$R$ is the radius of curvature of the AdS space and $\eta_{\mu \nu}={\rm diag }(1,-1,-1,-1)$. The $z$ (bulk) coordinate runs in the range $0 <z <+\infty$ (or, considering a UV cutoff, in the range $\varepsilon<z< +\infty$ with small positive $\varepsilon\to 0$). 
The model is characterized by a background dilaton field $\phi(z)$ which depends on the bulk coordinate and appears in the Lagrangian as  $e^{-\phi(z)}$ \cite{Karch:2006pv}. 
With the minimal choice $\phi(z)=c^2 z^2$ linear Regge trajectories for the spectra of light vector mesons \cite{Karch:2006pv}, light scalar mesons \cite{Colangelo:2008us} and scalar glueballs \cite{Colangelo:2007pt} are recovered. 
Conformal invariance in the ($z \to \infty$) IR is broken by the dimensionful constant $c$, and confinement is implemented. 
Other choices for the dilaton profile are described e.g. in Ref.~\cite{Chen:2022pgo}. The soft-wall is an example of holographic models; another model is the hard-wall, where $\phi=0$ and $z$ varying in a range up to a chosen IR value \cite{Erlich:2005qh}.

For $n_f$ light flavours the gauge fields $A_L^M(x,z)$ and $A_R^M(x,z)$ are introduced in the model, dual to the conserved left- and right-handed chiral quark current densities. 
They are defined as $A^M_{L(R)}(x,z)=A^{M,A}_{L(R)}(x,z) T^A$.
$T^A$ ($A=0,...,n_f^2-1$) are the $U(n_f)_{L(R)}$ generators, with $T^0=\frac{1}{2n_f} I_{n_f}$, where $I_{n_f}$ is the $n_f \times n_f$ identity matrix, and $\mathrm{Tr}[T^a T^b]=\delta^{ab}/2$ ($a,b=1,...,n_f^2-1$).
Vector and axial-vector fields are obtained by the combinations $V=(A_L+A_R)/2$ and $A=(A_L-A_R)/2$.
In the gauge $A_5^A=0$ the axial-vector fields can be written as the sum of the transverse $A_\perp^A$ and longitudinal $\varphi^A$ components: $A_\mu^A = A_{\perp\, \mu}^A + \partial_\mu \varphi^A$. 

A bifundamental bulk scalar field $X(x,z)$ is included in the model to introduce explicit and spontaneous breaking of the $U(n_f)_R\times U(n_f)_L$ invariance,
\begin{equation}
X(x,z)=e^{i\eta^A(x,z) T^A} X_0(z) \, e^{i\eta^A(x,z) T^A} . \label{scalarquark}
\end{equation}
$X(x,z)$ is dual to the $\bar q_R q_L(x)$ boundary operator. We consider the case $n_f=2+1$ with degenerate up and down quarks, and set $X_0(z)$, the dual of the boundary vev, of the form 
\begin{equation}\label{eq:X0}
X_0(z)=\sqrt{2} \, {\rm diag}(v_q(z),v_q(z),v_s(z)) \, .
\end{equation}
For the functions $v_{q(s)}(z)$ we assume the truncated near-boundary (low-$z$) expansion $\displaystyle v_{q(s)}(z)=\frac{m_{q(s)}}{R} z+\frac{\sigma_{q(s)}}{R} z^3$, with $m$ interpreted as the light quark mass $m_u=m_d=m_q$ and $m_s\neq m_q$, and $\sigma$ as the chiral condensate, according to the gauge/gravity dictionary. In our analysis we consider two possibilities: $\sigma_q=\sigma_s= \sigma$, and $\sigma_s= 0.8 \,\sigma_q$ as obtained by QCD sum rules \cite{Colangelo:2000dp}.

To face the $U(1)_A$ problem, another scalar field is included in the model, $Y(x,z)$, with modulus dual to the square of the gluon field strength $\alpha_s G^2$, and phase dual to $\alpha_s G \tilde G$, as proposed in \cite{Katz:2007tf}. The same term has been considered in \cite{Colangelo:2007pt} to compute the scalar glueball spectrum.
Written in the form
\begin{equation}
Y(x,z) = Y_0(z) \, e^{2ia(x,z)} , \label{scalargluon}
\end{equation}
$Y_0(z)$ is dual to the QCD gluon condensate. 
The main difference between the approach presented here and the hard-wall model of Ref.~\cite{Katz:2007tf} is in the transformation rules of fields under $U(1)_A$:
\begin{gather}
 \eta^0 \to \eta^0 -\alpha \\
 \varphi^0 \to \varphi^0 -\alpha \\
 a \to a - V_a \alpha \,. \label{eq:aU1}
 \end{gather}
In the present model $V_a(z)$ is a potential term, first introduced in \cite{Arean:2016hcs}, while in Ref.~\cite{Katz:2007tf} $V_a(z)=1$ is used. 
The potential generates a different interaction Lagrangian for the $X$ and $Y$ fields.
We generalize the expression of $V_a(z)$ in \cite{Arean:2016hcs} to the 2+1 flavour case assuming
\begin{equation}\label{eq:Va}
V_a(z)=e^{-(2v_q(z)^2+v_s(z)^2)/3}\,.
\end{equation}

The nonet of light pseudoscalar mesons together with the light pseudoscalar glueball are encoded in the phases in \eqref{scalarquark}, in 
the longitudinal modes of the axial-vector fields $\varphi^A$ and in the phase in \eqref{scalargluon}.

The holographic model is defined by the $5d$ action \cite{Giannuzzi:2021euy}
 \begin{equation}
S= \int \mathrm{d}^4 x \int_0^ {+\infty} \mathrm{d} z \,\mathcal{L} \,\, , \label{action}
\end{equation}
with 
\begin{equation}\label{eq:toymodel}
\mathcal{L}=\frac{1}{k}\sqrt{g}\, e^{-\phi(z)}\,\left\{ \mathrm{Tr}\left[ -\frac{1}{4g_5^2} (F_L^2+F_R^2) + |DX|^2 - m_X^2 |X|^2\right] +\frac{1}{2} \mathcal{K}_a \right\} .
\end{equation}
$g$ is the determinant of the metric, $F_{MN}$ is defined as $F_{MN}=\partial_M A_M-\partial_N A_M-i [A_M,A_N]$ for both the left and right gauge fields, 
the covariant derivative of the scalar field $DX$ is defined as 
$D X = \partial X -i A_L X +i X A_R$.
According to the gauge/gravity dictionary, the mass of the $X$ field is set to $m_X^2=-3/R^2$ from the dimension of the $\bar q_R q_L(x)$ boundary operator.
The interaction term $\mathcal{K}_a$ in \eqref{eq:toymodel}, proposed in \cite{Casero:2007ae}, has been introduced in the soft-wall model in \cite{Giannuzzi:2021euy} in the three flavour symmetric case, providing a solution to the $U(1)_A$ problem different from the one in \cite{Katz:2007tf}. 
Here we use the same kinetic term for the 2+1 flavour case, with potential from Eq.~\eqref{eq:Va}, assuming:
\begin{equation}\label{eq:kineticY}
 \mathcal{K}_a=\left | \partial_M Y_0(z) + 2 i Y_0 (z) \left (\partial_M a(x,z) - \eta^0(x,z) \partial_M V_a(z) - A_M^0(x,z) V_a(z)\right) \right |^2 \,.
\end{equation}
$Y_0(z)$ in Eq.~\eqref{scalargluon} is obtained from the equation of motion,
\begin{equation}\label{eq:Y0general}
Y_0(z)=\frac{y_0}{R}+\frac{2\, y_1}{R\, c^4} (e^{c^2 z^2} (-1 + c^2 z^2)+1) 
\end{equation}
and involves two integration constants $y_0$ and $y_1$. 
The first two terms in the near-boundary expansion of \eqref{eq:Y0general} are therefore a constant term and a term proportional to $z^4$. 
In the model in \cite{Katz:2007tf} only the first term is considered, while  in \cite{Leutgeb:2022lqw} the coefficient of the $z^4$ term with a logarithmic correction is included.
Eqs.~\eqref{eq:kineticY} and \eqref{eq:Y0general} characterize the holographic model \cite{Giannuzzi:2021euy}.

Let us focus on the term in the Lagrangian density (\ref{eq:toymodel}),(\ref{eq:kineticY}) comprising the fields $ \varphi^0, \varphi^8$, $\eta^0$, $\eta^8$ and $a$. In the gauge $A_5=0$ it reads:
\begin{eqnarray}\label{eq:final-Lagrangian}
 \mathcal{L}_{\eta,\eta^\prime} &=& \frac{R}{k} e^{-\phi(z)}\left[ \frac{1}{4 n_f g_5^2 z}\, (\partial_z \partial_\nu \varphi^0)^2 + \frac{1}{2g_5^2 z}\, (\partial_z \partial_\nu \varphi^8)^2 \right.\nonumber\\
 && -\frac{2 R^2\, (2 v_q^2+v_s^2)}{n_f^2 z^3} (\partial_z \eta^0)^2+\frac{2 R^2\, (2 v_q^2+v_s^2)}{n_f^2 z^3} (\partial_\nu \eta^0-\partial_\nu \varphi^0)^2\nonumber\\
 && -\frac{4\, R^2\, (v_q^2+2 v_s^2)}{n_f z^3} (\partial_z \eta^8)^2+\frac{4\, R^2\, (v_q^2+2 v_s^2)}{n_f z^3} (\partial_\nu \eta^8-\partial_\nu \varphi^8)^2 \nonumber\\
 && -\frac{8 (v_q^2- v_s^2)}{n_f^{3/2} z^3} ( \partial_z \eta^0) ( \partial_z \eta^8)+\frac{8 (v_q^2- v_s^2)}{n_f^{3/2} z^3} ( \partial_\nu\eta^0- \partial_\nu\varphi^0) ( \partial_\nu\eta^8 - \partial_\nu\varphi^{8}) \nonumber\\
 && \left.-\frac{2 R^2}{z^3} Y_0^2 (\partial_z a-\eta^0 \partial_z V_a)^2+\frac{2 R^2}{z^3}Y_0^2 (\partial_\nu a-V_a \partial_\nu \varphi^0)^2\right]\,.
\end{eqnarray}
From now on we set $R=1$. 
The $4D$ Fourier transforms of $\varphi^8(x,z)$, $\eta^8(x,z)$, $\varphi^0(x,z)$, $\eta^0(x,z)$ and $a(x,z)$ are obtained solving a system of five coupled equations of motions derived from  \eqref{eq:final-Lagrangian}: 
\begin{eqnarray}
 && \partial_z\left( \frac{e^{-\phi}}{2 n_f g_5^2 z} \partial_z \varphi^0 \right) +\frac{4 e^{-\phi} (2 v_q^2+v_s^2)}{n_f^2 z^3} ( \eta^0 -\varphi^0) +\frac{8 e^{-\phi} (v_q^2- v_s^2)}{n_f^{3/2} z^3} ( \eta^8 - \varphi^{8})\nonumber\\
 && +\frac{4 e^{-\phi}}{z^3}Y_0^2 V_a (a-V_a \varphi^0)=0 \label{sys1} 
\end{eqnarray}
\begin{eqnarray}
 && \partial_z\left( \frac{ e^{-\phi}}{g_5^2 z} \partial_z \varphi^8 \right) +\frac{8 e^{-\phi} (v_q^2+2 v_s^2)}{n_f z^3} (\eta^8- \varphi^8) +\frac{8 e^{-\phi} (v_q^2- v_s^2)}{n_f^{3/2} z^3} ( \eta^0- \varphi^0) =0 \label{sys2}
\end{eqnarray}
\begin{eqnarray}
 && \partial_z\left( \frac{4 e^{-\phi}}{z^3} Y_0^2 (\partial_z a-\eta^0 \partial_z V_a) \right) +\frac{4 e^{-\phi} q^2}{z^3}Y_0^2 (a-V_a \varphi^0) = 0 \label{sys3}
\end{eqnarray}
\begin{eqnarray}
 \frac{q^2 }{2 n_f g_5^2 z} \partial_z \varphi^0 - \frac{4 (2 v_q^2+v_s^2)}{n_f^2 z^3} ( \partial_z \eta^0 ) -\frac{8 (v_q^2- v_s^2)}{n_f^{3/2} z^3} ( \partial_z \eta^8) - \frac{4}{z^3} Y_0^2 V_a (\partial_z a-\eta^0 \partial_z V_a) =0 \label{sys4}
\end{eqnarray}
\begin{eqnarray}
 \frac{q^2}{g_5^2 z} \partial_z \varphi^8 - \frac{8 (v_q^2+2 v_s^2)}{n_f z^3} ( \partial_z \eta^8) - \frac{8 (v_q^2- v_s^2)}{n_f^{3/2} z^3} ( \partial_z \eta^0) =0\,. \label{sys5}
 \end{eqnarray}
The Dirichlet conditions $\eta^8=\varphi^8=\eta^0=\varphi^0=a=0$ at the boundary $z \to 0$, and the conditions $\partial_z \eta^8=\partial_z\varphi^8=\partial_z \eta^0=\partial_z\varphi^0= a=0$ in the IR $z\to\infty$ are imposed to obtain the masses $m_n^2=q^2$ and wavefunctions of the pseudoscalar states $\eta$, $\eta^\prime$ and $a$.
Notice that an IR Dirichlet boundary condition is also required for $a(z)$, since $a'(z)$  vanishes as $z\to \infty$ due to  $Y_0(z)$ in Eq.~\eqref{sys3}.
Each eigenvalue refers to a physical (eigen)state described by all the functions ($\varphi^0$, $\varphi^8$, $a$, $\eta^0$, $\eta^8$) with different weights.
In our notation for the fields, the superscript indicates the flavour content,  the subscript  the physical state (particle).
Hence, $(\varphi^8_\eta,\eta^8_\eta)$ describes the octet flavour content of $\eta$, $(\varphi^8_{\eta^\prime},\eta^8_{\eta^\prime})$  the octet flavour content of  $\eta^\prime$.
The normalization condition is imposed \cite{Erlich:2005qh}:
\begin{eqnarray}\label{eq:normalization}
 && \frac{R}{kg_5^2} \int_0^{+\infty} \mathrm{d}z \, e^{-\phi(z)} \left( \frac{(\partial_z \varphi^0)^2}{2 n_f z} + \frac{(\partial_z \varphi^8)^2}{z} + \frac{4 g_5^2 (2v_q^2+v_s^2) (\eta^0-\varphi^0)^2}{n_f^2 z^3} + \frac{8 g_5^2 (v_q^2+2v_s^2) (\eta^8-\varphi^8)^2}{n_f z^3}\right.\nonumber\\
 &&\left. + \frac{16 g_5^2 (v_q^2-v_s^2) (\eta^8-\varphi^8)(\eta^0-\varphi^0)}{n_f^{3/2} z^3} + \frac{4 g_5^2 Y_0^2}{z^3} (a-V_a \varphi^0)^2\right) = 1 \,.
\end{eqnarray}

$\pi^0$ is described by the terms in the Lagrangian density \eqref{eq:toymodel} comprising  $\varphi^3$ and $\eta^3$:
\begin{eqnarray}\label{eq:pion-Lagrangian}
    \mathcal{L}_{\pi^0} &=& \frac{R}{k} e^{-\phi(z)}\left[  \frac{1}{2g_5^2 z}\, (\partial_z \partial_\nu \varphi^3)^2 -\frac{4\, R^2\, v_q^2}{z^3} (\partial_z \eta^3)^2+\frac{4\, R^2\, v_q^2}{z^3} (\partial_\nu \eta^3-\partial_\nu \varphi^3)^2 \right]\,.
   \end{eqnarray}
The pion mass and wavefunction are obtained by  the  coupled equations
\begin{eqnarray}
    && \partial_z\left( \frac{ e^{-\phi}}{g_5^2 z} \partial_z \varphi^3 \right) +\frac{8 e^{-\phi} v_q^2}{z^3} (\eta^3- \varphi^3) =0 \label{eq:pionphi}
   \end{eqnarray}
\begin{eqnarray}
    \frac{q^2}{g_5^2 z} \partial_z \varphi^3 - \frac{8 v_q^2}{z^3} ( \partial_z \eta^3)  =0\,, \label{eq:pioneta}
    \end{eqnarray}
with boundary conditions $\eta^3=\varphi^3=0$ at $z \to 0$, and $\partial_z \eta^3=\partial_z\varphi^3=0$ in the  $z\to\infty$ IR region, and
 normalization  \cite{Erlich:2005qh}:
\begin{equation}\label{eq:normalizationpion}
 \frac{R}{kg_5^2} \int_0^{+\infty} \mathrm{d}z \, e^{-\phi(z)} \left( \frac{(\partial_z \varphi^3)^2}{z}  + \frac{8 g_5^2 v_q^2 (\eta^3-\varphi^3)^2}{z^3}\right) = 1 \,.
\end{equation}

The model is defined matching the two-point correlation functions of the vector and scalar quark currents to the perturbative QCD expressions, which fix $R/k=N_c/16 \pi^2$ and $g_5^2=3/4$ \cite{Colangelo:2008us}. 
The parameter $k$ has been introduced in \cite{Colangelo:2008us}  to simultaneously obtain the  coefficients of the leading terms in the high-$Q^2$ expansions of the two-point correlation functions of vector and scalar currents.
The scale $c$ in the dilaton field can be obtained from the $\rho$ meson mass, by the relation $m_n^2=4 c^2 (n+1)$ for the light vector meson spectrum \cite{Karch:2006pv}: $c=388$ MeV.
Deriving the on-shell action with respect to $m_q$ the relation $\langle \bar q q\rangle= -\frac{N_c}{2 \pi^2} \sigma $ is obtained, setting $\sigma$.

The following conditions allow to fix the numerical values of the other parameters.
In the chiral limit Eqs. \eqref{eq:pionphi}-\eqref{eq:pioneta} can be analytically solved and only depend on $\sigma$. Then, the pion decay constant can be analytically computed from
\begin{eqnarray}
    f_\pi&=&\left.\frac{R}{kg_5^2} e^{-\phi}\frac{\partial_z\varphi_\pi^3}{z}\right|_{z=0}\nonumber\\
    & =&  \left(\frac{R}{kg_5^2} \left(-c^2-\frac{(4 g_5 \sigma)^{2/3}}{\pi} \frac{-1+\pi \, \mathrm{Ai}(c^4/(4 g_5\sigma)^{4/3})\, \mathrm{Bi}'(c^4/(4 g_5\sigma)^{4/3})}{\mathrm{Ai}(c^4/(4 g_5\sigma)^{4/3})\, \mathrm{Bi}(c^4/(4 g_5\sigma)^{4/3})}\right)\right)^{1/2}\,,
\end{eqnarray}
where $\varphi_\pi^3$ is the normalized pion wavefunction, and $\mathrm{Ai}$ and $\mathrm{Bi}$ are the Airy functions.
From $f_\pi=92.4$ MeV, we fix $\sigma=0.149$ GeV$^3$, which gives $\langle \bar q q\rangle = (- 0.283~\mathrm{GeV})^3$. 
Using as input the pion and $\eta$ masses ($m_{\pi^0}=134.9$ MeV and $m_\eta=547.8$ MeV) we fix $m_q = 3.47~\mathrm{MeV}$ and $m_s = 101.5~\mathrm{MeV}$.
The parameter $y_0=\alpha_s/(\pi \sqrt{N_c})$ in Eq.~\eqref{eq:Y0general} is obtained from the high-$Q^2$ expansion of the two-point correlation function of the pseudoscalar $G\tilde G$ operator \cite{Giannuzzi:2021euy}. Here $\alpha_s$ is not running, and its value can be fixed, e.g.,  by the pseudoscalar glueball decay constants. 
Since they are not precisely known, we set  $y_0=1/(\pi \sqrt{N_c})$.  Then $y_1=0.041/\sqrt{N_c}$ GeV$^4$ comes from the pure-gauge topological susceptibility \cite{Giannuzzi:2021euy}, obtaining a successful prediction for the $\eta^\prime$ mass.
In the hard-wall model in \cite{Katz:2007tf} the parameter $y_1$ is neglected, an assumption modified in \cite{Leutgeb:2022lqw}. In the present model neglecting $y_1$ and considering a constant function $Y_0$ leads to a vanishing topological susceptibility \cite{Giannuzzi:2021euy}. 
After fixing the parameters of the model, we obtain the prediction $m_{\eta^\prime}=957.7$ MeV for the $\eta^\prime$ mass.
From the two-point correlation function of the pion interpolating current at finite quark mass we obtain   $f_\pi=93.1$ MeV.
Setting instead $\sigma_s=0.8 \, \sigma_q$ \cite{Colangelo:2000dp}, and fitting again the strange quark mass $m_s=111.5$ MeV from the $\eta$ mass, we find $m_{\eta^\prime}=983.8$ MeV.

\begin{figure}[ht]
 \centering
 \includegraphics[width=0.5\textwidth]{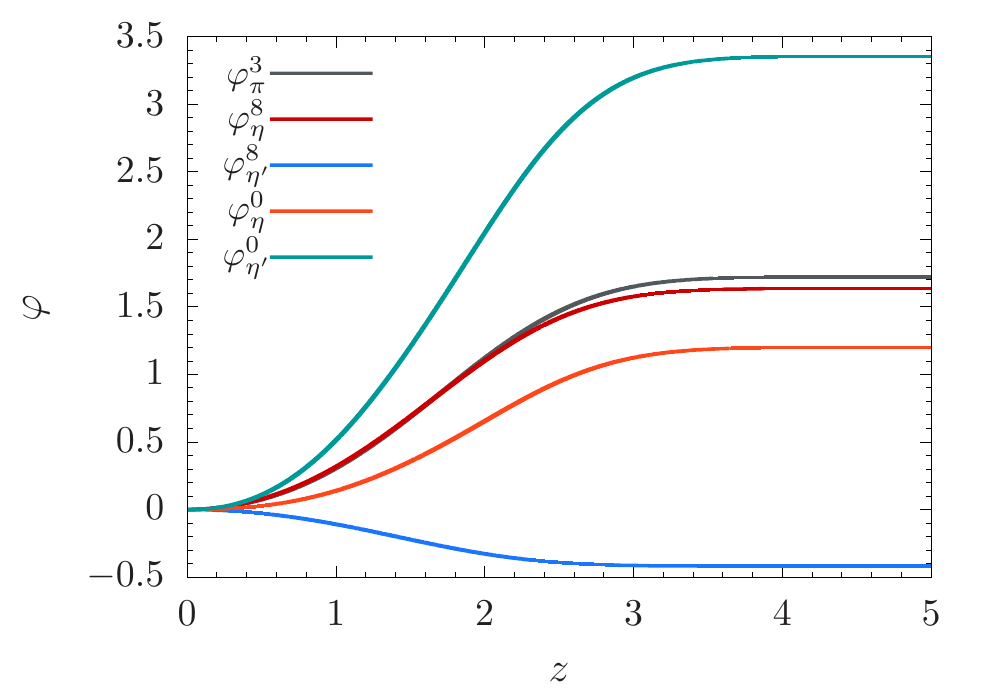}
 \caption{Normalized functions $\varphi^A_P(z)$ for $\pi^0, \eta$ and $\eta^\prime$.}
 \label{fig:wavefunctions}
\end{figure}

The system (\ref{sys1})-(\ref{sys5}) of differential equations is solved using both the  Mathematica  NDSolve function and an implementation of the Runge-Kutta method in C language. The eigenvalues are found solving the boundary value problem by the Newton's method.
The normalized functions $\varphi_P^{A}(z)$ of $\pi^0$, $\eta$ and $\eta^\prime$ are depicted in Fig.~\ref{fig:wavefunctions}. As expected, for $\eta$ the octet component $\varphi^8_\eta$ is dominant over the singlet $\varphi^0_\eta$, while the opposite holds for  $\eta^\prime$.

\section{Two-photon transition form factor of light pseudoscalar mesons}\label{sec:TFF}

We now consider the transition  form factor (TFF)  $F_{\pi^0 \gamma^* \gamma^*}$,   defined as  
\be
i \int \mathrm{d}^4 x\,  e^{i q \cdot x} \langle 0| T[ j_\mu(x) j_\nu(0)] | \pi^0 (p) \rangle \, = \, \epsilon_{\mu \nu \rho \sigma} \, q^\rho \, p^\sigma \, F_{\pi^0 \gamma^* \gamma^*}(q^2, (p-q)^2) \,\, ,
\ee
and analogously for the other mesons. $j_\mu(x)$ is the hadronic electromagnetic current.
In the holographic approach $F_{P \gamma^* \gamma^*}$ can be computed incorporating in the model the chiral anomaly, adding to the action \eqref{action} the $5D$ Chern-Simons (CS) term
\begin{equation}
S_{CS}=S_{CS}^L-S_{CS}^R . \label{CSaction}
\end{equation}
The two terms in \eqref{CSaction} read
\begin{equation}\label{eq:CS2}
 S_{CS}^{L(R)} = \frac{N_c}{24\pi^2} \int \, \mathrm{Tr}\left[\mathcal{A} \mathcal{F}^2 -\frac{i}{2} \mathcal{A}^3 \mathcal{F}-\frac{1}{10} \mathcal{A}^5\right]^{L(R)} ,
\end{equation}
with $\mathcal{A}=A_M \mathrm{d}x^M$, $\mathcal{F}=(\partial_A A_B)\, \mathrm{d}x^A\wedge \mathrm{d}x^B$, $A$ being the left (right) $A_{L(R)}$ gauge field.
The action \eqref{CSaction} has been considered for the analysis of the $AVV^*$ vertex function  \cite{Colangelo:2011xk}.\footnote{ Other computations of $F_{\pi^0 \gamma^* \gamma^*}$ in holographic models are in Refs. \cite{Grigoryan:2007wn,Kwee:2007dd,Grigoryan:2008up,Zuo:2011sk,Brodsky:2011xx}.}
 
 The transition form factor of the pseudoscalar meson $P$ to two photons with spacelike virtualities $Q_1^2$ and $Q_2^2$ can be obtained  from  \eqref{CSaction},\eqref{eq:CS2} \cite{Leutgeb:2021mpu}:
\begin{equation}
 F_{P\gamma^*\gamma^*}(Q_1^2,Q_2^2)=-\frac{N_c}{2\pi^2} x^A \, K^A_P(Q_1^2,Q_2^2) \,.
\end{equation}
This expression  involves the overlap integral
\begin{equation}
 K^A_P(Q_1^2,Q_2^2)=-\int_0^\infty \mathrm{d}z\, J(Q_1^2,z) \, J(Q_2^2,z) \partial_z \varphi_P^A(z)\, \label{eq:KA}
\end{equation}
and the factor $x^A=\mathrm{Tr}[T^A Q_{em}^2]$. $Q_{em}$ is the light quarks electromagnetic charge matrix, so that $x^3=1/6$, $x^8=1/(6\sqrt{3})$, $x^0=1/9$.
In the integral \eqref{eq:KA},  $J(Q^2,z)$  is  the bulk-to-boundary propagator of the vector field for spacelike $Q^2$, which in the soft-wall model reads
\begin{equation}
 J(Q^2,z)=\frac{Q^2}{4c^2} \, \Gamma\left(\frac{Q^2}{4c^2}\right) \, U\left(\frac{Q^2}{4c^2},0,c^2z^2\right)\,,
\end{equation}
with $U$ the Tricomi confluent hypergeometric function \cite{Karch:2006pv,Colangelo:2011xk}. 
At $Q^2=0$, the bulk-to-boundary propagator $J$ is constant: $J(0,z)=1$.

In the chiral limit with the normalization \eqref{eq:normalizationpion} one has $\varphi^3_{\pi^0}=1/f_\pi$ at $z\to\infty$,  and the pion form factor to two real photons coincides with the result from the chiral anomaly \cite{Abidin:2009aj,Leutgeb:2021mpu}:
\begin{equation}
F_{\pi^0\gamma\gamma}=\frac{N_c}{12\pi^2 f_\pi}\,,
\end{equation}
corresponding to $F_{\pi^0\gamma\gamma}=0.2739$ GeV$^{-1}$.
Notice that in the soft-wall model additional boundary terms in the Chern-Simons action are not needed to compute $F_{P\gamma^*\gamma^*}$, since  the equations of motion force $\varphi^A(z) = \eta^A(z)$ for $z\to\infty$. 

 For the first radial excitation of the pion $\pi^0(2S)$ we obtain $m_{\pi^0(2S)}= 2.087$ GeV and $F_{\pi^0(2S)\gamma\gamma} = -0.2952$ GeV$^{-1}$.
For $\eta$ and $\eta^\prime$ we find $F_{\eta\gamma\gamma}=0.2777$ GeV$^{-1}$ and $F_{\eta^\prime\gamma\gamma}=0.3175$ GeV$^{-1}$,
 results  collected in Table~\ref{tab:results}. The corresponding $\eta$ and $\eta^\prime$ two-photon decay widths are
 $\Gamma(\eta \to \gamma \gamma)=0.531$ keV and
  $\Gamma(\eta^\prime \to \gamma \gamma)=3.71$ keV,
  compared to the measurements
   $\Gamma(\eta \to \gamma \gamma)_{exp}=0.516 \pm 0.020$ keV and 
  $\Gamma(\eta^\prime \to \gamma \gamma)_{exp}=4.32 \pm 0.15$ keV \cite{ParticleDataGroup:2022pth}.
  Setting $\sigma_s=0.8 \, \sigma_q$ we find $F_{\eta\gamma\gamma}=0.2808$ GeV$^{-1}$ and $F_{\eta^\prime\gamma\gamma}=0.3039$ GeV$^{-1}$ (Table~\ref{tab:results}).
  
  The next isoscalar pseudoscalar ground state meson, resulting from the $\eta_8$, $\eta_0$ and pseudoscalar glueball mixing, turns out to have a low mass, i.e. 1.14 GeV, and a sizeable two-photon width. Therefore, its contribution to $a_\mu$ is sizeable as well,  $a_\mu=5.07\times 10^{-11}$ as obtained using the expressions in the next section.  On the other hand,  pseudoscalar glueballs and excited $\eta$ mesons are estimated to give small contributions to $a_\mu$, see e.g. Ref.~\cite{Hechenberger:2023ljn} in which these states have been considered in a Witten-Sakai-Sugimoto model and Ref.~\cite{Leutgeb:2022lqw} for the hard-wall model. In the soft-wall model, even in pure-gauge theory, the  pseudoscalar glueballs are much lighter  than in other models or in  lattice QCD \cite{Giannuzzi:2021euy}, a theoretical uncertainty inducing us not to include the results for the next isoscalar pseudoscalar  state in Table \ref{tab:results}.

 \begin{table}[ht]
 \begin{center}
 \begin{tabular}{ ccccc } 
 \hline
 & $\pi^0$ & $\pi^0{(2S)}$ & $\eta$ & $\eta^\prime$ \\ 
 \hline
 mass (MeV) & 134.9 & 2087 & 547.8 (547.7) & 957.7 (983.8) \\
 $F_{P\gamma\gamma}$ $(\mbox{GeV}^{-1})$ & 0.2739 & -0.2952 & 0.2777 (0.2808) & 0.3175 (0.3039) \\ 
 $a_\mu^{HLbL}$ $ \times 10^{11}$ & 75.2 & 1.68 & 21.2 (22.1) & 12.3 (10.5) \\ 
 \hline
 \end{tabular}
 \end{center}
 \caption{\small Mass, transition form factor $F_{P \gamma\gamma}$ at $Q^2_1=Q^2_2=0$, and HLbL contribution to $a_\mu$ of the pseudoscalar mesons $P=\pi^0, \pi^0(2S), \eta, \eta^\prime$ for 
 $\sigma_q=\sigma_s=\sigma$. The results obtained for $\sigma_s= 0.8 \, \sigma_q$ are displayed in brackets.}
 \label{tab:results}
\end{table}

\begin{figure}[h!]
 \centering
 \includegraphics[width=0.5\textwidth]{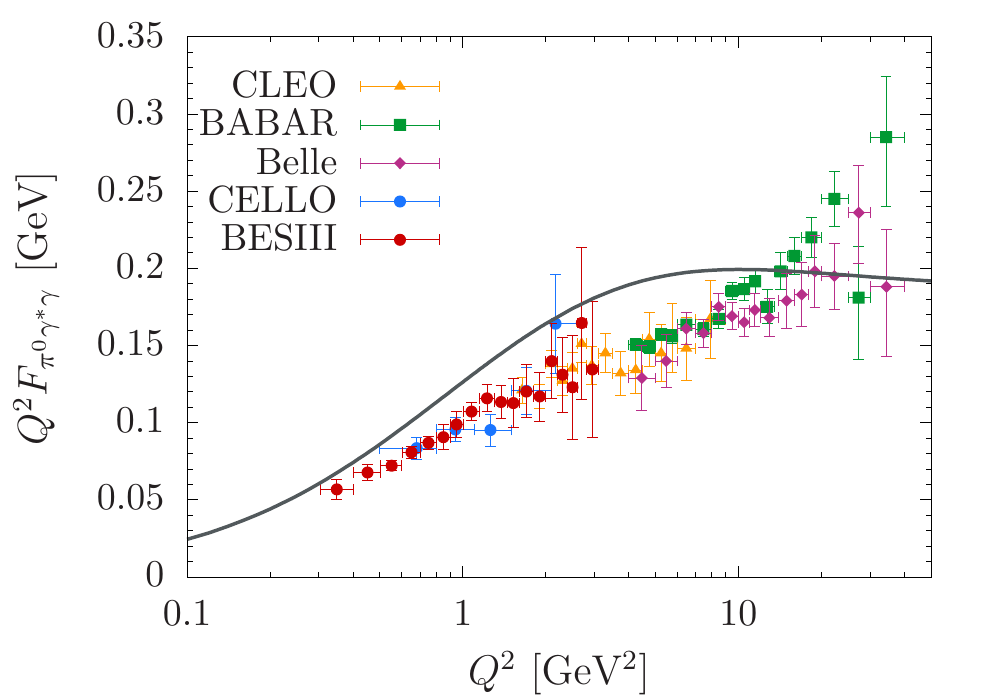}
 \caption{\small $\pi^0 \to \gamma^* \gamma$ transition form factor (continuous line), and experimental results by the Collaborations CELLO \cite{CELLO:1990klc}, CLEO \cite{CLEO:1997fho}, BaBar \cite{BaBar:2009rrj}, Belle \cite{Belle:2012wwz} and BES III (preliminary) \cite{Redmer:2018uew}.}
 \label{fig:TFF-real-pion-log}
\end{figure}

The $\pi^0$ TFF for one real and one virtual photon is depicted in Fig.~\ref{fig:TFF-real-pion-log}, together  with the  experimental results.
For the chosen expression of the dilaton and numerical values of the parameters, the result displays a sharper $Q^2$ dependence than the  measurements in the range up to $Q^2\sim 8$ GeV$^2$.  At larger $Q^2$  the asymptotic Brodsky-Lepage behaviour  is recovered, with $Q^2  F_{\pi^0 \gamma^*\gamma}(Q^2,0)  \to 2 f_\pi$  \cite{Lepage:1980fj}.
The increase found by  the BaBar Collaboration  for $Q^2 > 10$ GeV$^2$ is not reproduced.

\begin{figure}[t]
 \centering
 \includegraphics[width=0.5\textwidth]{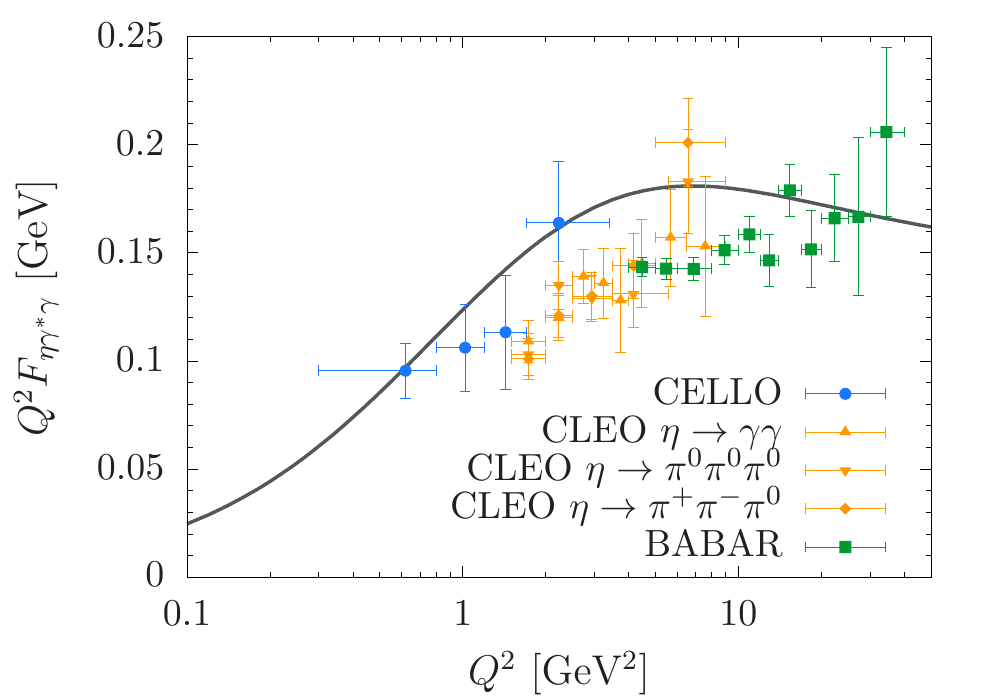}
 \caption{\small $\eta \to \gamma^* \gamma$ transition form factor (continuous line) and measurements by CELLO \cite{CELLO:1990klc}, CLEO \cite{CLEO:1997fho} and BaBar Collaboration \cite{BaBar:2009rrj}.}
 \label{fig:TFF-real-eta-log}
\end{figure}

\begin{figure}[h]
 \centering
 \includegraphics[width=0.5\textwidth]{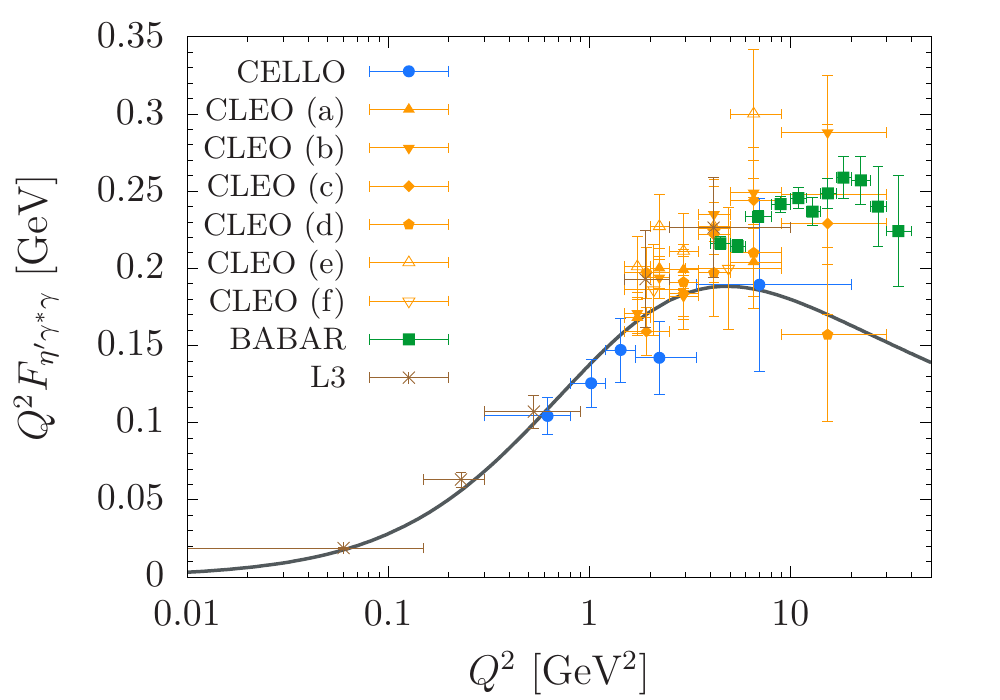}
 \caption{\small $\eta^\prime \to \gamma^* \gamma$ transition form factor (continuous line) and measurements by the Collaborations CELLO \cite{CELLO:1990klc}, CLEO \cite{CLEO:1997fho}, BaBar \cite{BaBar:2009rrj} and L3 \cite{L3:1997ocz}. In the legend, CLEO (a) refers to the process $\eta^\prime\to\pi^+\pi^-\gamma$, CLEO (b) to $\eta^\prime\to\pi^+\pi^-\eta(\to\gamma\gamma)$, CLEO (c) to $\eta^\prime\to\pi^+\pi^-\eta(\to\pi^+\pi^-\pi^0)$, CLEO (d) to $\eta^\prime\to\pi^+\pi^-\eta(\to 3\pi^0)$, CLEO (e) to $\eta^\prime\to\pi^0\pi^0\eta(\to\gamma\gamma)$, CLEO (f) to $\eta^\prime\to\pi^0\pi^0\eta(\to 3\pi^0)$.}
 \label{fig:TFF-real-etap-log}
\end{figure}

The TFF for $\eta$ and $\eta^\prime$, for one real and one virtual photon,  are shown in Figs.~\ref{fig:TFF-real-eta-log} and \ref{fig:TFF-real-etap-log}. 
As for the pion,  the $\eta$ form factor slightly exceeds the measurements in the $Q^2$ bins where the errors are small. For $\eta^\prime$ there is agreement up to 
$Q^2 \simeq 8$ GeV$^2$,  above this value the result deviates from measurements.
At large $Q^2$, $Q^2 F_{\eta\gamma^*\gamma}$ and $Q^2 F_{\eta^\prime\gamma^*\gamma}$ approach a constant value. By a polynomial fit in $1/Q^2$, we find that the asymptotic values are 0.15 GeV for $\eta$ and 0.1 GeV for  $\eta^\prime$.

\begin{figure}[ht]
 \centering
 \includegraphics[width=0.5\textwidth]{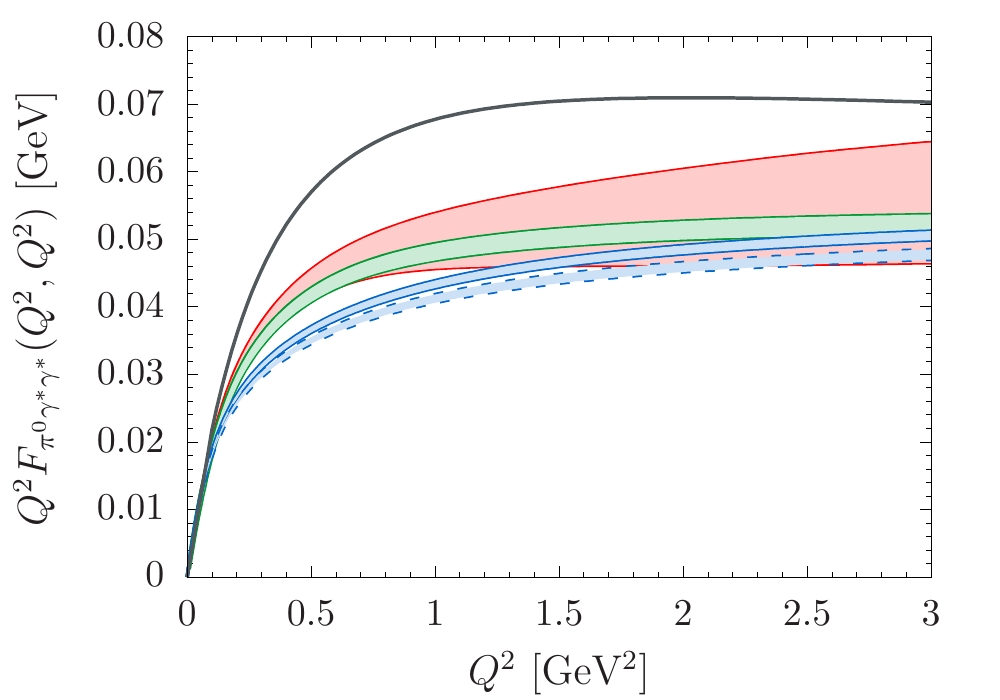}
 \caption{ \small Double virtual pion TFF with $Q^2=Q_1^2=Q_2^2$  (black line), and as obtained in Ref. \cite{Hoferichter:2018kwz} (red band), in Ref. \cite{Gerardin:2019vio} (green band) and in  Ref. \cite{Danilkin:2019mhd} (blue bands) for the model parameter $\Lambda^2=0.611\pm 0.005$ GeV$^2$ (plain blue lines) and $\Lambda^2=0.574\pm 0.007$ GeV$^2$ (dashed blue  lines).}
 \label{fig:TFFvirtualpion}
\end{figure}

\begin{figure}[ht]
 \centering
 \includegraphics[width=0.5\textwidth]{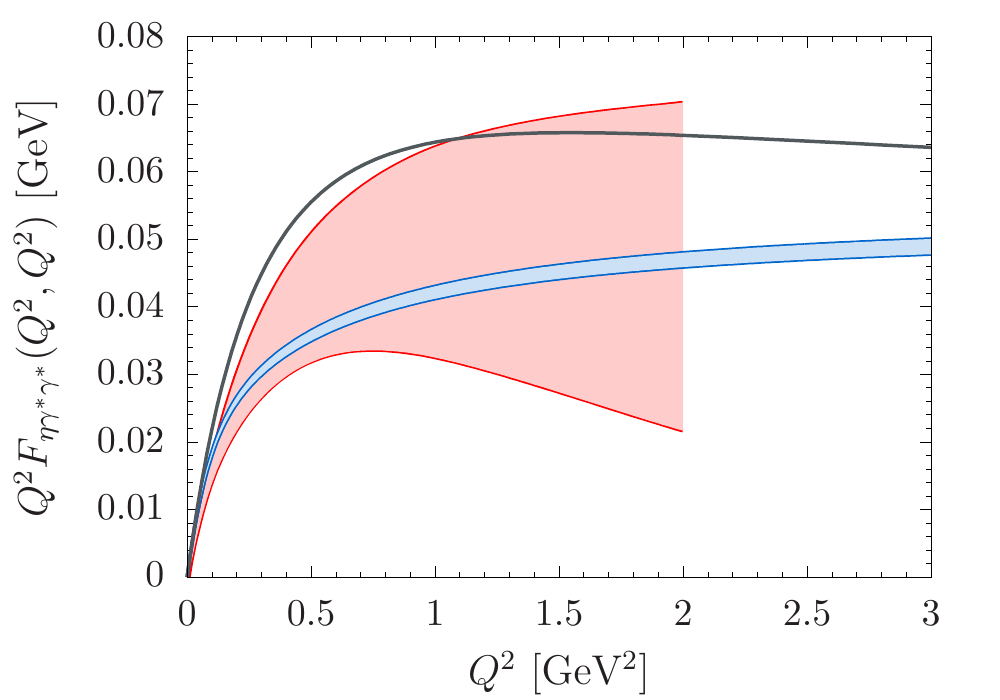}
 \caption{ \small Double virtual $\eta$ TFF with $Q_1^2=Q_2^2=Q^2$ (black line) and by the calculations in \cite{Alexandrou:2022qyf} (red band) and  \cite{Danilkin:2019mhd} (blue band).}
 \label{fig:TFFvirtualeta}
\end{figure}

\begin{figure}[ht]
 \centering
 \includegraphics[width=0.5\textwidth]{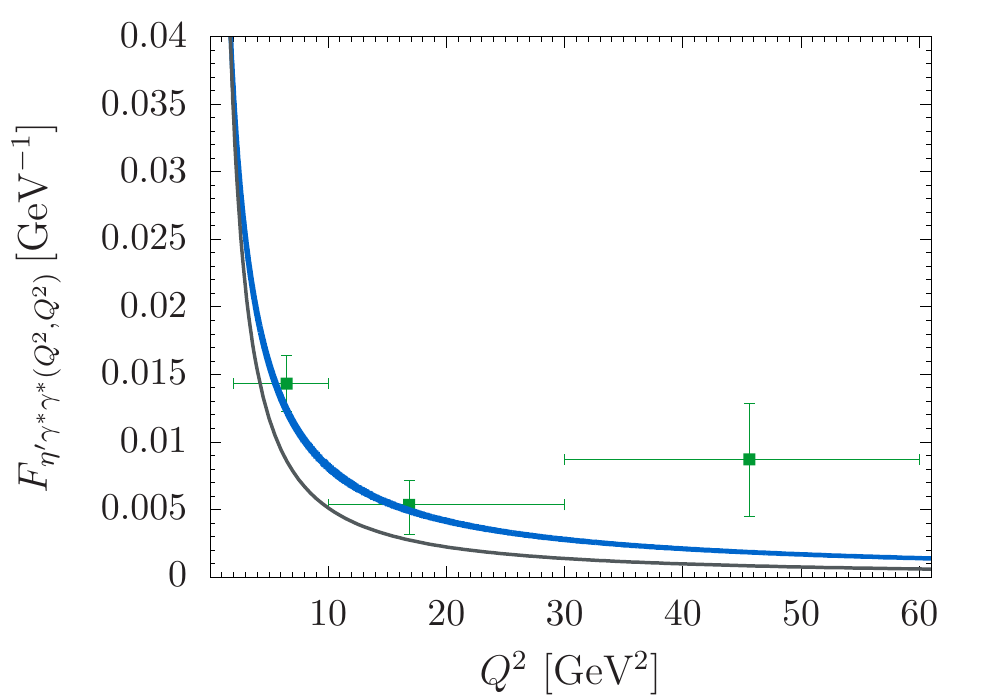}
 \caption{ \small Double virtual $\eta^\prime$ TFF with $Q_1^2=Q_2^2=Q^2$ (black line) and by the calculation in \cite{Danilkin:2019mhd} (blue band). The green points are the BaBar Collaboration measurements \cite{BaBar:2018zpn}.}
 \label{fig:TFFvirtualetap}
\end{figure}

The TFF for two virtual photons, in the symmetric $Q_1^2=Q_2^2$ configuration, are depicted in Figs.~\ref{fig:TFFvirtualpion}-\ref{fig:TFFvirtualetap}. 
The only available  measurements from the BaBar Collaboration  concern  $\eta^\prime$ \cite{BaBar:2018zpn} for which there is agreement in the first two bins of momentum. 
For $\pi^0$ and $\eta$ the curves are compared to lattice QCD results \cite{Gerardin:2019vio,Alexandrou:2022qyf}, the dispersive results \cite{Hoferichter:2018kwz} and the parameterization of Ref.~\cite{Danilkin:2019mhd}. 
Also in the double virtual case, the obtained curves for $Q^2 F_{P\gamma^*\gamma^*}$ are higher than other determinations.

The various transition form factors have been recently computed in an updated holographic  hard-wall model, with results generally in better agreement with experiment  \cite{Leutgeb:2022lqw}. 
One has to stress that the soft-wall model is characterized by a minimal set of parameters, not fine-tuned, allowing to describe a broad phenomenology.
To investigate the sensitivity to a single parameter, we have changed the value of the scale $c$ which is not fixed by the pseudoscalar meson masses. Reducing $c$ results in a better agreement of $F_{\pi^0 \gamma^* \gamma}$ with measurements at low and intermediate $Q^2$, however the agreement with the meson spectroscopy is degraded: as an example, a smaller $c$ corresponds to a lower mass for the $\rho$ meson.
In Ref. \cite{Brodsky:2011xx}  $F_{\pi^0 \gamma^* \gamma}$  has been computed in a light-front soft-wall holographic model in the chiral limit, and it displays a monotonic behaviour. The difference with our result is not due to the different value of the quark mass, which produces a tiny effect, but  to the different pion wavefunction obtained in the two approaches.

The pseudoscalar  meson form factors  could be improved considering a more complex dilaton profile or a dynamical dilaton, generally renouncing to  analytical solutions. 
It has been pointed out that in the soft-wall model, even using a dynamical dilaton field, the light-flavour hadron spectra and the pion form factor cannot be easily reconciled: a possible solution is to include an anomalous correction in  the mass of the  scalar field $X$ \cite{Chen:2022pgo}.

\section{ Pole contribution of pseudoscalar mesons  to $a_\mu^{HLbL}$}\label{sec:g-2}
The pseudoscalar meson two-photon transition form factors determine the pole contribution to $a_\mu^{HLbL}$ \cite{Colangelo:2015ama}.
The pole contribution of $\pi^0$, $\eta$ and $\eta^\prime$ to $a_\mu^{HLbL}$ can be computed using the expression \cite{Nyffeler:2016gnb}
\begin{eqnarray}\label{eq:amuLbL}
 a_\mu^{HLbL} &=& \left(\frac{\alpha}{\pi}\right)^3\int_0^\infty \mathrm{d}Q_1 \int_0^\infty \mathrm{d}Q_2 \int_{-1}^{1} \mathrm{d}\tau\, \left( w_1(Q_1,Q_2,\tau) F_{P\gamma^*\gamma^*}(Q_1^2,(Q_1+Q_2)^2) F_{P\gamma^*\gamma^*}(Q_2^2,0)\right.\nonumber\\
 && \left. + w_2(Q_1,Q_2,\tau) F_{P\gamma^*\gamma^*}(Q_1^2,Q_2^2) F_{P\gamma^*\gamma^*}((Q_1+Q_2)^2,0)\right)\,,
\end{eqnarray}
with $Q_{1,2}=\sqrt{Q^2_{1,2}}$. 
The functions $w_1$ and $w_2$ are in the  Appendix A of Ref.~\cite{Nyffeler:2016gnb}. 
The holographic expression  \eqref{eq:amuLbL} is a $5$-dimensional integral, with two integrals in the bulk coordinate $z$ (one for each form factor in Eq.~\eqref{eq:KA}), an integral in the angular variable $\tau$ and two integrals in  $Q_{1,2}$. 
We used the Monte Carlo VEGAS algorithm with integration limits $-1+\delta\leqslant \tau \leqslant 1-\delta$, $Q_{min}\leqslant Q_{1,2} \leqslant Q_{max}$, $\varepsilon\leqslant z \leqslant z_{max}$. The values $Q_{min}=10^{-4}$ GeV, $Q_{max}=30$ GeV, $\delta=10^{-5}$, $\varepsilon=10^{-4}$ GeV$^{-1}$, and $z_{max}\sim 5$ GeV$^{-1}$ for $\pi^0$, $\eta$ and $\eta^\prime$, and $z_{max}\sim 5.6$ GeV$^{-1}$ for $\pi^0(2S)$ have been chosen. Stability against variation of such values has been checked. The results are collected in Table \ref{tab:results}.

The HLbL pole  contribution of  different mesons  to $a_\mu$ has been computed by various holographic models \cite{Hong:2009zw,Cappiello:2010uy,Cappiello:2019hwh,Leutgeb:2019zpq,Leutgeb:2021mpu}. Scalar mesons have been considered in \cite{Cappiello:2021vzi}, axial-vector mesons   in \cite{Leutgeb:2019gbz}, scalar and pseudoscalar glueballs in \cite{Hechenberger:2023ljn}.
 $a_\mu^{HLbL}$ has been computed for the pion, $\eta$ and $\eta^\prime$ in the hard-wall model with finite quark masses considering a finite number of vector meson modes contributing to the $\gamma^*\gamma^*$ transition form factor \cite{Hong:2009zw}.
The  lightest pseudoscalar mesons contribution has been computed in the soft-wall model in the chiral limit  \cite{Leutgeb:2019zpq}. The $\eta$ and $\eta^\prime$ contributions for finite quark masses in a hard-wall model with a solution of the $U(1)_A$ problem are evaluated in \cite{Leutgeb:2022lqw}.

Due to the larger TFF the pion pole contribution is higher than the one obtained in Ref. \cite{Leutgeb:2022lqw}.
For  $\pi^0(2S)$ we obtain $a_\mu = 1.68\times 10^{-11}$, a value larger than the one obtained in the hard-wall model in \cite{Leutgeb:2022lqw}, but similar to the one obtained in \cite{Leutgeb:2021mpu} by a modified version of the hard-wall model in which the 5$d$ mass of the scalar $X$ field deviates from $m_5^2=-3$.
In \cite{Cappiello:2019hwh} two sets of parameters have been chosen in a hard-wall model in which spontaneous chiral symmetry breaking is implemented by different boundary conditions for left and right fields.
With the first set the $\rho$ meson mass is reproduced, while the second  set is used to best fit the pion decay constant.
With set 1 (set 2) the obtained pion TFF and  $a_\mu^{HLbL}$ are smaller than (similar to) those determined here.

We find that the contribution of the first radial excitation  of the pion is  $2.2 \%$ of the $\pi^0$ pole, those of 
$\eta$ and $\eta^\prime$  are $28.2 \%$ and $16.4 \%$ of the pion pole, respectively,  see Table \ref{tab:results}.
 The excitations of $\eta$ and $\eta^\prime$ have not been considered, their contribution is expected to be smaller than $\pi^0(2S)$. 
 In all cases the error associated to the numerical computation is controlled and tiny. 

The results obtained setting  $\sigma_s=0.8 \sigma_q$ are also collected in Table~\ref{tab:results}. The variation with respect to the case $\sigma_s= \sigma_q$ can be considered as a theoretical uncertainty.


The White Paper \cite{Aoyama:2020ynm} reports for the $\pi^0$, $\eta$ and $\eta^\prime$ HLbL pole contributions:
\bea
    a_\mu^{\pi^0\mbox{-pole}} (\mbox{disp})&=&63.0^{+2.7}_{-2.1} \times 10^{-11}\, \nn \\
    a_\mu^{\pi^0\mbox{-pole}} (\mbox{CA})&=&63.6(2.7) \times 10^{-11}\, \nn \\
    a_\mu^{\pi^0\mbox{-pole}} (\mbox{lattice})&=&62.3(2.3) \times 10^{-11}\,\\
    a_\mu^{\eta\mbox{-pole}} (\mbox{CA})&=&16.3(1.4) \times 10^{-11}\, \nn \\
    a_\mu^{\eta^\prime\mbox{-pole}} (\mbox{CA})&=&14.5(1.9) \times 10^{-11}\, , \nn
\eea
obtained from dispersive analyses (disp),  Canterbury approximants (CA) and lattice QCD. 
 New lattice determinations  are in \cite{Alexandrou:2022qyf,Verplanke:2021gat}. 
The comparison with the results  in Table~\ref{tab:results} reflects the comparison between the computed and measured TFFs:  a higher value is obtained for $\pi^0$, compatible results for  $\eta$ and  $\eta^\prime$.

We have investigated how the results depend on the value of $\alpha_s$ appearing in the parameter $y_0$ in \eqref{eq:Y0general}. 
Setting $\alpha_s=0.5$, from the topological susceptibility we fix $y_1=0.048$ GeV$^4$ \cite{Giannuzzi:2021euy}, and fitting the strange quark mass from the $\eta$ mass, we find  $m_{\eta^\prime}=1.012$ GeV. In this case the TFFs with one virtual photon keep qualitatively the same shapes, but get higher values than in the $\alpha_s=1$ case, with a larger effect for $\eta^\prime$. 
The pole contributions to $a_{\mu}$ increase: $a_{\mu}^{\eta}=21.9\times 10^{-11}$ and $a_{\mu}^{\eta^\prime}=19.2\times 10^{-11}$. 
Using $\alpha_s=0.5$, a smaller topological susceptibility is needed in order to obtain the $\eta^\prime$ mass in agreement with the experiments, namely $\chi_{PG}\sim (178\mbox{ MeV})^4$ instead of $\chi_{PG}\sim (191\mbox{ MeV})^4$ used in our study.

\section{Conclusions}\label{sec:conclusions}
We have computed the two-photon transition form factors of  $\pi^0$, $\eta$ and $\eta^\prime$ mesons in a minimal soft-wall holographic model of QCD  modified to face the $U(1)_A$ problem.  The  comparison  with the experimental data in different ranges of photon virtualities shows that,
despite its simplicity, the model captures the main phenomenological features  employing a small number of parameters.
To achieve a better  agreement with experimental data, a more elaborated model must be developed involving, e.g.,  anomalous dimension effects for various  operators.

The sum of the $\pi^0$, $\eta$ and $\eta^\prime$ HLbL pole contributions  to the muon anomalous magnetic moment is $108.7\times10^{-11}$,  compared to  $94.3(5.3)\times10^{-11}$ quoted in   \cite{Aoyama:2020ynm}. Including  $\pi^0(2S)$  the sum increases to  $110.4\times10^{-11}$. 
The estimated contribution of $\eta^{\prime\prime}$ increases  the sum by a further $5.07\times 10^{-11}$, with sizeable uncertainties related to the description of pseudoscalar glueballs in the soft-wall model.

Improving the theoretical determinations of  the hadronic contribution to the muon anomalous magnetic moment is a difficult task requiring the use of many different methods.
The comparison between the outcome of the various methods is important to assess the theoretical error affecting such determinations. In our study we have considered the pole contributions of the lightest pseudoscalar mesons using the soft-wall holographic model of QCD.
Other resonances with higher spin and/or heavier mass can affect the muon anomalous magnetic moment. Although they  are expected to provide a smaller contribution, it would be interesting to precisely determine their effect using our model. In particular, as discussed in \cite{Aoyama:2020ynm}, the axial-vector mesons are responsible for a sizeable portion of the theoretical uncertainty of the HLbL contribution, since the estimates from different hadronic models vary quite strongly. For axial-vector mesons
the hard-wall holographic models predict larger values than most hadronic models \cite{Leutgeb:2019gbz}, hence it would be interesting to see whether large contributions are also obtained in the soft-wall model.

\section*{Acknowledgements}
This study has been carried out within the INFN project (Iniziativa Specifica) QFT-HEP.

\bibliographystyle{JHEP}
\bibliography{gm2.bib}

\end{document}